\documentclass[a4paper,12pt]{article}  

\usepackage{graphicx}
\usepackage{amssymb}
\usepackage{amsmath}

\title{A phase-integral perspective on $\alpha$-decay}

\date{\today}

\author{Giampiero Esposito ORCID: 0000-0001-5930-8366 \\
Istituto Nazionale di Fisica Nucleare, Sezione di
Napoli, \\ 
Complesso Universitario di Monte S. Angelo, \\ 
Via Cintia Edificio 6, 80126 Napoli, Italy}

\begin{document}
\maketitle

\begin{abstract}
This paper applies the phase-integral method to the stationary
theory of $\alpha$-decay. The rigorous form of the connection
formulae, and their one-directional nature
that was not widely known in the physical literature, are applied. 
The condition for obtaining $s$-wave metastable states affects
the stationary state at large distance from the nucleus, which is
dominated by the cosine of the phase integral minus ${\pi \over 4}$. 
Accurate predictions for the lowest $s$-wave metastable state
and mean life of the radioactive nucleus are obtained in the
case of Uranium. The final part of the paper describes the
phase-integral algorithm for evaluating stationary states 
by means of a suitable choice of freely specifiable base function.
Within this framework, an original approximate formula for the
phase integrand with arbitrary values of the angular momentum
quantum number is obtained.
\end{abstract}

\section{Introduction}
\setcounter{equation}{0}

The detailed investigation of $\alpha$-decay is a topic that 
leads to a thorough understanding of the application of quantum
mechanics to atomic and nuclear physics, 
since it is necessary to have a good knowledge of metastable states
\cite{B1,B2} and of the Jeffreys-Wentzel-Kramers-Brillouin 
(hereafter, JWKB) method 
applied to the Schr\"{o}dinger equation for stationary states
\cite{B2,B3,B4}. In particular, in the description of the JWKB
method, the introductory textbooks on quantum mechanics fail,
even nowadays, to present the remarkable results obtained in
Refs. \cite{B3,B4}, which are written in a very clear and pedagogical style.

Our paper lies precisely within this framework. In section $2$
we outline some relevant features of the phase-integral method
and of the associated derivation of connection formulae. In 
section $3$ we consider the basic equations for the elementary 
stationary theory of $\alpha$-decay. Section $4$
develops a more accurate model of the stationary
theory in section 3, and improves the theoretical estimate
of the lowest $s$-wave metastable state. The mean life of the
radioactive nucleus is evaluated in section $5$ for Uranium. 
Section $6$ describes the phase-integral algorithm for evaluating
stationary states by means of a suitable choice of freely
specifiable base function \cite{B3,B4}, while
concluding remarks are presented in section $7$. 

\section{Phase integral method and connection formulae}
\setcounter{equation}{0}

Both in one-dimensional problems and in the case of central potentials
in three-dimensional Euclidean space, the Schr\"{o}dinger equation
for stationary states leads eventually to a second-order ordinary
differential equation having the form \cite{B1,B2,B3,B4}
\begin{equation}
\left[{d^{2}\over dz^{2}}+R(z)\right]\psi(z)=0, \; \; \;
R(z)={2m \over \hbar^{2}}(E-V(z)),
\label{(2.1)}
\end{equation}
where $V(z)$ is either the potential in one spatial dimension, or
an effective potential that includes also the effects of angular
momentum. The notation $z$ for the independent variable means that
one can study Eq. (2.1) in the complex field, restricting attention
to real values of $z$, denoted by $x$, only at a later stage. 
In the phase-integral method, one looks for two linearly independent,
exact solutions of Eq. (2.1) in the form
\begin{equation}
\psi(z)=A(z) e^{\pm i w(z)}.
\label{(2.2)}
\end{equation}
Since the Wronskian of two linearly independent solutions of Eq. (2.1)
is a non-vanishing constant, while the Wronskian of the functions (2.2)
is $-2iA^{2}{dw \over dz}$, for consistency one finds
\begin{equation}
A={{\rm constant}\over \sqrt{dw \over dz}}.
\label{(2.3)}
\end{equation}
One can therefore write (up to a multiplicative constant)
\begin{equation}
\psi(z)={1 \over \sqrt{dw \over dz}}e^{\pm i w(z)}
={1 \over \sqrt{q(z)}} e^{\pm i \int^{z} q(\zeta)d\zeta},
\label{(2.4)}
\end{equation}
where $w(z) \equiv \int^{z}q(\zeta)d\zeta$ is said to be the
{\it phase integral}, while $q(z)$ is the {\it phase integrand}
\cite{B3,B4}.

In quantum mechanical problems, we shall agree to call 
{\it classically forbidden} the open interval of the independent
variable where the energy $E$ of the particle is strictly less
than the potential $V$: $E < V$.
Conversely, if $E>V$, we shall talk of {\it classically allowed}
region. For the former, let $x_{1}$ be an internal
point, where the stationary state takes the exact form 
(hereafter, the independent variable is always real)
\begin{equation}
\psi(x_{1})=
c(x_{1}){1 \over \left | \sqrt{q(x_{1})} \right | }
e^{|w(x_{1})|}
+d(x_{1}){1 \over \left | \sqrt{q(x_{1})} \right | }
e^{-|w(x_{1})|},
\label{(2.5)}
\end{equation}
$c$ and $d$ being real-valued functions. For the latter, let
$x_{2}$ be an internal point, where the stationary state
takes the exact form
\begin{equation}
\psi(x_{2})=
a(x_{2}){1 \over \left | \sqrt{q(x_{2})} \right | }
e^{i|w(x_{2})|}
+b(x_{2}){1 \over \left | \sqrt{q(x_{2})} \right | }
e^{-i|w(x_{2})|}.
\label{(2.6)}
\end{equation}
From the detailed theory in section $2.4$ of Ref. \cite{B4},
one knows that (hereafter, the star denotes complex conjugation)
\begin{equation}
a(x_{2})=\left \{ {1 \over 2 \alpha} c(x_{1})
+i \alpha \Bigr[\gamma c(x_{1})-d(x_{1})\Bigr] \right \}
e^{i \left({\pi \over 4}-\beta \right)},
\label{(2.7)}
\end{equation}
\begin{equation}
b(x_{2})=a^{*}(x_{2}).
\label{(2.8)}
\end{equation}
The exact expressions of the $\alpha,\beta,\gamma$ parameters are known
(see Ref. \cite{B4} and our appendix A)
and they are not particularly enlightening. However, upon setting
\begin{equation}
\chi(q(x)) \equiv q^{-{3 \over 2}}{d^{2}\over dx^{2}}
q^{-{1 \over 2}}+{R(x) \over q^{2}}-1,
\label{(2.9)}
\end{equation}
\begin{equation}
\mu(x,x_{0}) \equiv \left | \int_{x_{0}}^{x} 
\mid \chi(q(x')) \; q(x') \mid  dx' \right | ,
\label{(2.10)}
\end{equation}
if $\mu <<1$, one can use the approximate formulae \cite{B4}
\begin{equation}
\alpha \sim 1+ {\rm O}(\mu), \; \;
B \sim {\rm O}(\mu), \; \;
\gamma \sim {\rm O}(\mu)e^{2 |w(x_{1}|}.
\label{(2.11)}
\end{equation}
If $\mu$ is much bigger than $e^{-2|w(x_{1})|}$, one then finds from
the third of Eqs. (2.11)
\begin{equation}
\gamma \sim {\rm O}(\mu)e^{2|w(x_{1})|} >> 1.
\label{(2.12)}
\end{equation}
Since $\gamma$ is unknown and in general much bigger than $1$,
one can obtain approximate formulae for $a(x_{2})$ and $b(x_{2})$ 
only when 
\begin{equation}
\left | \gamma c(x_{1}) \right | <<
\left | d(x_{1}) \right | .
\label{(2.13)}
\end{equation}
By virtue of the condition (2.12), the majorization (2.13) provides
\begin{equation}
\left | {c(x_{1}) \over d(x_{1})} \right | \leq 
e^{-2 |w(x_{1}|}.
\label{(2.14)}
\end{equation}
If the right-hand side of (2.14) is much smaller than $1$, the
exact formula (2.7) yields the remarkable approximate formula
\begin{eqnarray}
a(x_{2}) & \approx & \left[{1 \over 2} c(x_{1})-i d(x_{1}) \right]
e^{i \left({\pi \over 4}-\mu \right)} 
\nonumber \\
& \approx & d(x_{1}) \left[{1 \over 2}
{c(x_{1}) \over d(x_{1})}-i \right] e^{i {\pi \over 4}}
\approx -i d(x_{1}) e^{i {\pi \over 4}}
\nonumber \\
& = & d(x_{1}) e^{i \left({\pi \over 4}-{\pi \over 2}\right)}
=d(x_{1}) e^{-i{\pi \over 4}},
\label{(2.15)}
\end{eqnarray}
while 
\begin{equation}
b(x_{2})=a^{*}(x_{2}) \approx d(x_{1}) e^{i{\pi \over 4}}.
\label{(2.16)}
\end{equation}
The exact formula (2.6) leads therefore to the approximate formula
\begin{eqnarray}
\psi(x_{2}) & \approx & 
d(x_{1})e^{-i{\pi \over 4}} 
\left | q^{-{1 \over 2}}(x_{2})
\right | e^{i |w(x_{2})|}
+d(x_{1}) e^{i {\pi \over 4}} 
\left | q^{-{1 \over 2}}(x_{2})
\right | e^{-i |w(x_{2})|}
\nonumber \\
&=& 2 d(x_{1}) \left | q^{-{1 \over 2}}(x_{2}) \right |
\cos \left[|w(x_{2})|-{\pi \over 4} \right].
\label{(2.17)}
\end{eqnarray}
From Eqs. (2.5) and (2.17) one gets therefore the connection formula
\begin{equation}
c \left | q^{-{1 \over 2}}(x) \right | e^{|w(x)|}
+d \left | q^{-{1 \over 2}}(x) \right | e^{-|w(x)|}
\longrightarrow
2d \left | q^{-{1 \over 2}}(x) \right | \cos
\left[ |w(x)| -{\pi \over 4} \right],
\label{(2.18)}
\end{equation}
which holds provided that the condition (2.13) is fulfilled.
In the literature, the case $c=0,d=1$ is often considered for 
simplicity \cite{B4}. Remarkably, the connection formula (2.18)
is one-directional. The work in Ref. \cite{B4} proves indeed that,
if one is first given a stationary state in the classically
allowed region having the form
$$
\psi(x_{2})=2 \left | q^{-{1 \over 2}}(x_{2}) \right |
\cos \left[|w(x_{2})| -{\pi \over 4} \right ],
$$
the stationary state in the classically forbidden region does
not reduce to the left-hand side of Eq. (2.18) with $c=0$
and $d=1$. This property is so important for our analysis that
we prove it in appendix B, so that our paper becomes completely
self-contained.

\section{Elementary stationary theory of $\alpha$-decay}
\setcounter{equation}{0}

As is well known, experiments in which a sufficiently large number of
$\alpha$-particles enter a chamber with a thin window and are collected
show that $\alpha$-rays correspond to positively charged particles whose
charge-to-mass ratio is the one of doubly ionized helium atoms:
${\rm He}^{++}$. Their identification with ${\rm He}^{++}$ is made
possible because, when a gas of $\alpha$-particles produces light, it 
displays precisely the spectroscopic lines of ${\rm He}^{++}$. In the
stationary theory of $\alpha$-emission, one regards the $\alpha$-particle
as being pre-existent in the radioactive nucleus. Such a radioactive
nucleus is therefore viewed as a metastable state\footnote{Recall from
the theory of resonance scattering \cite{B1,B2} that there exists
a metastable state corresponding to a trapping of the particle
in the region where the potential makes its effect manifest.}  
consisting of the $\alpha$-particle and the residual nucleus. 
The force acting on the $\alpha$-particle is the joint effect of a
short-range nuclear interaction and a long-range Coulomb repulsion. 
The long-range component is described by a potential
$2(Z-2){e_{0}^{2}\over r}$, and the potential is assumed to obey
the defining law
\begin{equation}
V(r)=-U_{0} \; \; {\rm if} \; \; r \in ]0,b[,
\label{(3.1)}
\end{equation}
\begin{equation}
V(r)=2(Z-2){e_{0}^{2}\over r} \; \; r \in ]b,\infty[.
\label{(3.2)}
\end{equation}
The $s$-wave metastable states can be found by solving the equation
(cf. Eq. (2.1))
\begin{equation}
\left[{d^{2}\over dr^{2}}+{2m \over \hbar^{2}}(E-V(r)) \right]
y(r)=0,
\label{(3.3)}
\end{equation}
in the three open intervals
\begin{equation}
I_{1}=]0,b[, \; \; I_{2}=]b,r_{1}[, \; \;
I_{3}=]r_{1},\infty[,
\label{(3.4)}
\end{equation}
where $r_{1}$ is the value of $r$ for which the energy of the
$\alpha$-particle equals the Coulomb term, i.e.
\begin{equation}
r_{1}=2(Z-2){e_{0}^{2}\over E}.
\label{(3.5)}
\end{equation}
The interval $I_{2}$ corresponds to values of the energy $E<V$, 
while the interval $I_{3}$ pertains to values of the energy $E>V$.
On defining
\begin{equation}
p_{0} \equiv \sqrt{2m_{\alpha}(E+U_{0})}, \; \;
{\bar p}(r) \equiv \sqrt{2m_{\alpha}(V(r)-E)}, \; \;
p(r) \equiv \sqrt{2m_{\alpha}(E-V(r))},
\label{(3.6)}
\end{equation}
which are appropriate for $I_{1},I_{2},I_{3}$, respectively,
one can write the solutions of Eq. (3.3) within such
intervals (see comments after Eq. (3.13)) in the form \cite{B2}
\begin{eqnarray}
\; & \; & y_{1}(r)= {C \over \sqrt{p_{0}}} 
\sin \left({p_{0}r \over \hbar}\right)
\nonumber \\
&=& {C \over \sqrt{p_{0}}} \left \{
A_{1} \sin \left[{p_{0}(b-r)\over \hbar}-{\pi \over 4}\right]
+A_{2} \cos \left[{p_{0}(b-r)\over \hbar}-{\pi \over 4}\right]
\right \},
\label{(3.7)}
\end{eqnarray}
\begin{equation}
y_{2}(r)={C \over \sqrt{{\bar p}(r)}} \left \{
A_{3} {\rm exp}\left[{1 \over \hbar} \int_{b}^{r}
{\bar p}(r')dr' \right]
+A_{4} {\rm exp}\left[-{1 \over \hbar} \int_{b}^{r}
{\bar p}(r')dr' \right] \right \},
\label{(3.8)}
\end{equation}
\begin{eqnarray}
y_{3}(r)&=& {C \over \sqrt{p(r)}} \left \{
A_{5} \cos \left[{1 \over \hbar} \int_{r_{1}}^{r}
p(r')dr' -{\pi \over 4} \right] \right .
\nonumber \\
&+& \left . A_{6} \sin \left[{1 \over \hbar} \int_{r_{1}}^{r}
p(r')dr' -{\pi \over 4} \right] \right \},
\label{(3.9)}
\end{eqnarray}
where \cite{B2}
\begin{equation}
A_{1}=-\cos \left({p_{0}b\over \hbar}-{\pi \over 4}\right),
\; \;  
A_{2}= \sin \left({p_{0}b\over \hbar}-{\pi \over 4}\right),
\label{(3.10)}
\end{equation}
\begin{equation}
A_{3}=-A_{1}, \; \;
A_{4}={1 \over 2}A_{2},
\label{(3.11)}
\end{equation}
and, considering the parameter 
\begin{equation}
\theta \equiv {\rm exp} \left[-{1 \over \hbar}
\int_{b}^{r_{1}} {\bar p}(r)dr \right] << 1,
\label{(3.12)}
\end{equation}
one finds the last two coefficients in the form
\begin{equation}
A_{5}={2A_{3}\over \theta}, \; \; 
A_{6}=-A_{4} \theta.
\label{(3.13)}
\end{equation}
It should be stressed that only Eq. (3.7) provides an exact solution,
in the open interval $I_{1}$, whereas Eqs. (3.8) and (3.9) provide 
approximate solutions in the open intervals $I_{2}$ and $I_{3}$,
respectively. The coefficients $A_{3}$ and $A_{4}$ of Eq. (3.8) are 
obtained in Ref. \cite{B2} from a connection recipe, but not 
from the continuity condition of stationary states and their first
derivative, unlike the work in Ref. \cite{Holstein}.

The work in Ref. \cite{B2} points out that, for metastable states
to occur, one has to maximize the derivative of the phase shift with
respect to the energy, and this implies in turn that the ratio of
values of the stationary state inside and outside the potential well
must be maximized. For this purpose, the authors of Ref. \cite{B2}
set to zero $A_{3}$ (and hence $A_{1}$ and $A_{5}$), finding therefore,
for the lowest $s$-wave metastable state,
\begin{equation}
{p_{0}b \over \hbar}-{\pi \over 4}={\pi \over 2} 
\Longrightarrow p_{0}={3 \over 4} {\pi \hbar \over b}.
\label{(3.14)}
\end{equation}
The coefficient of the $\sin$ function in (3.9) is then found to be
$-{\theta \over 2}$, from Eqs. (3.13) and (3.14). When the condition
(3.13) is fulfilled, the derivative of the phase shift takes the
approximate form \cite{B2}
\begin{equation}
{d \over dp}\delta \approx \pi \int_{0}^{b}
|y_{1}(r)|^{2}dr.
\label{(3.15)}
\end{equation}
In order to evaluate the integral on the right-hand side of (3.15),
one has to evaluate the $C$ coefficient in the formulae for stationary
states in the three intervals. For this purpose, one looks first at
the interval $I_{3}$, where (see Eq. (3.9))
\begin{eqnarray}
{1 \over \hbar} \int_{r_{1}}^{r}p(r')dr'
&=& {\sqrt{2 m_{\alpha}E} \over \hbar}
\int_{r_{1}}^{r} \left(1-{2(Z-2)e_{0}^{2} \over
Er'}\right)^{1 \over 2}dr'
\nonumber \\
& \approx & {pr \over \hbar}-m_{\alpha}
{2(Z-2)e_{0}^{2}\over \hbar p} \log \left ({2pr \over \hbar}\right) + ... .
\label{(3.16)}
\end{eqnarray}
Thus, at very large values of $r$, Ref. \cite{B2} finds
\begin{eqnarray}
\; & \; & y_{3}(r) \approx - {C \theta \over 2 \sqrt{p(r)}}
\sin \left[{1 \over \hbar}
\int_{r_{1}}^{r} p(r')dr'-{\pi \over 4} \right]
\nonumber \\
& \approx & {C \theta \over 2 \sqrt{p(r)}} 
\cos \left[{pr \over \hbar}-m_{\alpha}
{2(Z-2)e_{0}^{2}\over \hbar p}
\log \left({2pr \over \hbar}\right)+ \vartheta \right],
\label{(3.17)}
\end{eqnarray}
where the explicit form of $\vartheta$ is here inessential,
but we can say that it is linearly related to the 
phase shift \cite{B2}.
On the other hand, from the general analysis of Coulomb type
potentials, one knows that \cite{B2}
\begin{equation}
y_{3}(r) \approx \sqrt{2 \over \pi \hbar} 
\cos \left[{pr \over \hbar}-m_{\alpha}
{2(Z-2)e_{0}^{2}\over \hbar p}
\log \left({2pr \over \hbar}\right)+ \vartheta \right],
\label{(3.18)}
\end{equation}
and hence by comparison of Eqs. (3.17) and (3.18) one finds
\begin{equation}
C=\sqrt{2 \over \pi \hbar} {2 \sqrt{p} \over \theta}.
\label{(3.19)}
\end{equation}
This implies in turn that
\begin{equation}
y_{1}(r)=\sqrt{8 \over \pi \hbar}
\sqrt{p \over p_{0}}{1 \over \theta}
\sin \left({p_{0}r \over \hbar}\right),
\label{(3.20)}
\end{equation}
and therefore Eq. (3.15) yields
\begin{eqnarray}
{d \over dp}\delta & \approx & {8 \over \hbar \theta^{2}}
{p \over p_{0}} \int_{0}^{b} \sin^{2}
\left({p_{0}r \over \hbar}\right)dr
\nonumber \\
&=& {8 \over \hbar \theta^{2}} {p \over p_{0}}
\left[{b \over 2}-{\hbar \over 4 p_{0}}
\sin \left({2 p_{0}b \over \hbar}\right) \right]
\nonumber \\
& \approx  & {4b \over \hbar \theta^{2}}.
\label{(3.21)}
\end{eqnarray}
The mean life of the radioactive nucleus is then
\begin{equation}
\tau = {\hbar \over 2} {d \over dE} \delta
\approx {2 m_{\alpha} \over p} {1 \over \theta^{2}}.
\label{(3.22)}
\end{equation}
Bearing in mind that the parameter $\theta$ defined in Eq. 
(3.12) is the exponential of minus the integral
$$
{1 \over \hbar} \int_{b}^{r_{1}}
\sqrt{2 m_{\alpha} \left[2(Z-2){e_{0}^{2}\over r}
-E \right]} dr,
$$
the stationary theory studied so far yields therefore
the prediction
\begin{eqnarray}
\log(\tau) &=& \log \left({2b m_{\alpha} \over p}\right)
-\log \theta^{2} 
\nonumber \\
& \approx & \log(\tau_{0}) +2(Z-2)\pi {e_{0}^{2}\over \hbar}
\sqrt{2 m_{\alpha}\over E} 
\nonumber \\
&+& {8 \over \hbar} \sqrt{(Z-2)e_{0}^{2}m_{\alpha}b}.
\label{(3.23)}
\end{eqnarray}

\section{A more accurate model}
\setcounter{equation}{0}

The careful reader might have noticed that the first line of
Eq. (3.17) is at odds with the connection formula 
(2.18), whose left-hand side corresponds neatly to Eq. (3.8).
In section $3$ we stressed indeed that Eq. (3.8) is not an exact solution 
of the stationary Schr\"{o}dinger equation in the interval $I_{2}$,
because the coefficients $A_{3}=-A_{1}$ and $A_{4}={1 \over 2}A_{2}$
are obtained \cite{B4} from the connection recipe
\begin{equation}
\cos \left[|w(x)|-{\pi \over 4}\right] \rightarrow
{1 \over 2}e^{-|w(x)|}, \; \;
\sin \left[|w(x)|-{\pi \over 4} \right] \rightarrow 
-e^{|w(x)|},
\label{(4.1)}
\end{equation}
which contradicts the connection formulae in Refs. \cite{B3,B4}. On the
other hand, the use of two consecutive connection formulae may be
questionable as well. More precisely, on passing from the interval
$I_{1}$ to the interval $I_{2}$, the work in section $3.12$ of Ref.
\cite{B4} would suggest using, instead of Eq. (4.1), the one-directional
connection formula
\begin{equation}
{a \over |\sqrt{q(x)}|}e^{i|w(x)|}+{b \over |\sqrt{q(x)}|}e^{-i|w(x)|}
\rightarrow \left[a \; e^{-i{\pi \over 4}}
+b \; e^{i{\pi \over 4}}\right]
{e^{|w(x)|}\over |\sqrt{q(x)}|},
\label{(4.2)}
\end{equation}
which is valid when the absolute value
$\left | a \; e^{-i{\pi \over 4}}+b \; e^{i{\pi \over 4}} \right |$
is not too small compared to $|a|+|b|$ \cite{B4}. However, if in
the interval $I_{2}$ only the increasing exponential 
$e^{|w(x)|}$ survives, the connection formula (2.18) would tell
us that we should expect a vanishing stationary state in the
interval $I_{3}$. But this conclusion would be incorrect, as is
shown from Eq. (3.18), which does not rely upon any form of
connection formula. The deeper underlying reason might be, that the
rigorous connection formulae in Ref. \cite{B4} hold for adjacent
intervals, but their repeated use for a sequence of adjacent intervals
requires further work. For this reason, and inspired in part by the
work in Ref. \cite{Holstein}, we consider hereafter the following method.

In the interval $I_{1}$, we write simply the exact solution of the
$s$-wave stationary Schr\"{o}dinger equation in the form displayed
on the first line of Eq. (3.7). In the interval $I_{2}$, we look
for a solution in the form (3.8), but with values of $A_{3}$ and
$A_{4}$ not given by Eq. (3.11). We impose instead the continuity
conditions for stationary state and its first derivative, which
hold whenever the potential has a finite discontinuity
\cite{B2,Espo1,Espo2}. Hence we require that
\begin{equation}
\lim_{r \to b}y_{1}(r)=\lim_{r \to b}y_{2}(r),
\label{(4.3)}
\end{equation}
\begin{equation}
\lim_{r \to b}y_{1}'(r)=\lim_{r \to b}y_{2}'(r).
\label{(4.4)}
\end{equation}
Equations (4.3)-(4.4) are solved by (cf. Eq. (9) in Ref. \cite{Holstein})
\begin{equation}
A_{3}={1 \over 2} \left \{ \sqrt{p_{0}\over {\bar p}(b)} 
\cos \left({p_{0}b \over \hbar}\right)
+\sqrt{{\bar p}(b)\over p_{0}} \left[1+{\hbar \over 2}
{{\bar p}'(b)\over ({\bar p}(b))^{2}} \right]
\sin \left({p_{0}b \over \hbar}\right) \right \},
\label{(4.5)}
\end{equation}
\begin{equation}
A_{4}={1 \over 2} \left \{ - \sqrt{p_{0}\over {\bar p}(b)} 
\cos \left({p_{0}b \over \hbar}\right)
+\sqrt{{\bar p}(b)\over p_{0}} \left[1-{\hbar \over 2}
{{\bar p}'(b)\over ({\bar p}(b))^{2}} \right]
\sin \left({p_{0}b \over \hbar}\right) \right \}.
\label{(4.6)}
\end{equation}
At this stage, if we follow the physical requirement of Ref.
\cite{B4} and our section $3$ for obtaining $s$-wave metastable
states, i.e., that the coefficient $A_{3}$ should vanish, 
we get the equation
\begin{equation}
\left[1+{\hbar \over 2}
{{\bar p}'(b)\over ({\bar p}(b))^{2}} \right]
\tan \left({p_{0}b \over \hbar}\right)
=-{p_{0}\over {\bar p}(b)}.
\label{(4.7)}
\end{equation}
For example, in the case of Uranium \cite{Holstein}, the right-hand 
side of Eq. (4.7) equals $-{9 \over 50}$, and bearing in mind that
$$
1 >> {\hbar \over 2} {{\bar p}'(b) \over ({\bar p}(b))^{2}},
$$
the approximate root of Eq. (5.7) is equal to 
\begin{equation}
{p_{0}b \over \hbar} \equiv \rho \approx 2.963,
\label{(4.8)}
\end{equation}
whereas the value (3.14) for ${p_{0}b \over \hbar}$ is
approximately equal to $2.356$. We find therefore, for the
energy of the lowest $s$-wave metastable state,
\begin{equation}
E={(p_{0})^{2}\over 2 m_{\alpha}}-U_{0},
\label{(4.9)}
\end{equation}
with $p_{0}$ given by Eq. (4.8). The work in Ref. \cite{Holstein}
sets instead to zero the right-hand side of Eq. (4.7), which 
is not sufficiently accurate, at least in the case of Uranium.

At this stage, if we define
\begin{equation}
\nu \equiv {p_{0}\over {\bar p}(b)},
\label{(4.10)}
\end{equation}
we find from Eq. (4.6) a good approximation for $A_{4}$ in the form
\begin{equation}
A_{4} \approx {1 \over 2} \left[ -\sqrt{\nu}
\cos (\rho)+{1 \over \sqrt{\nu}}
\sin(\rho)\right],
\label{(4.11)}
\end{equation}
where $\rho$ solves the equation that ensures the vanishing 
of $A_{3}$:
\begin{equation}
\tan(\rho)=-\nu,
\label{(4.12)}
\end{equation}
which implies ($\cos(\rho)$ is negative
since $\rho$ is close to $\pi$ by virtue of Eq. (4.8))
\begin{equation}
\cos(\rho)=-{1 \over \sqrt{1+\nu^{2}}}, \; \;
\sin(\rho)={1 \over \sqrt{1+{1 \over \nu^{2}}}}.
\label{(4.13)}
\end{equation}
Hence we obtain
\begin{equation}
2A_{4}={2 \over \sqrt{\nu + {1 \over \nu}}}=f(\nu),
\label{(4.14)}
\end{equation}
where for Uranium, exploiting again the value of 
$\nu={9 \over 50}$ from Ref. \cite{Holstein}, we find
\begin{equation}
\sqrt{\nu + {1 \over \nu}}=\sqrt{2581 \over 450}
\approx 2.3949.
\label{(4.15)}
\end{equation}

\section{Mean life of the radioactive nucleus}
\setcounter{equation}{0}

By virtue of the connection formula (2.18), which
can be used because the condition (2.13) is fulfilled
having set $A_{3}=0$ in section $4$, we can now
write the stationary state in the interval $I_{3}$ in
the approximate form
\begin{equation}
y_{3}(r) \approx 2A_{4}\theta {C \over \sqrt{p(r)}} 
\cos \left[{1 \over \hbar} \int_{r_{1}}^{r} p(r')dr'
-{\pi \over 4} \right].
\label{(5.1)}
\end{equation}
If we require that such a function should take 
the form (3.18) at large $r$, we find by
comparison that, up to a sign,
\begin{equation}
C=\sqrt{2 \over \pi \hbar} {\sqrt{p} \over \theta}
{1 \over 2A_{4}}
=\sqrt{2 \over \pi \hbar} {2 \sqrt{p}\over \theta}
{1 \over 2 f(\nu)}.
\label{(5.2)}
\end{equation}
By comparison of Eqs. (5.2) and (3.19), and bearing in
mind Eq. (3.22), our prediction for the
mean life of the radioactive nucleus reads as 
\begin{equation}
\log(\tau) = \log \left({2bm_{\alpha}\over p}\right)
-2 \log(\theta)-2 \log(2 f(\nu)).
\label{(5.3)}
\end{equation}
whereas the work in Ref. \cite{B2} obtains (see Eq. (3.23))
\begin{equation}
\log(\tau)=\log \left({2bm_{\alpha}\over p}\right)
-2 \log(\theta).
\label{(5.4)}
\end{equation}
In light of Eq. (4.15), the difference between our result
(5.3) and the theoretical prediction (5.4) is a
constant factor which, for Uranium, equals $-1.025$.

\section{Beyond $s$-wave}
\setcounter{equation}{0}

In the investigation of bigger values of angular momentum 
quantum number, it may be appropriate to exploit a further
refined version of nuclear potential, and also the potentialities
of the phase-integral method with unspecified base function
\cite{B4}, a concept that we are going to define shortly.

The models of current interest study the relative motion of 
the $\alpha$-particle and daughter nucleus in a central 
potential $U(r)$ built as follows. On considering the decay
of nuclei surrounded by electrons, the $\alpha$-particle moves
in the central potential \cite{DZ}
\begin{equation}
U(r)=U_{n}(r)+U_{C}(r),
\label{(6.1)}
\end{equation}
where $U_{n}(r)$ is the nuclear potential well, and 
$U_{C}(r)$ is the effective Coulomb potential. At small 
distances, when the $\alpha$-particle moves inside the nucleus
or under the barrier, the Coulomb contribution can be 
approximated (up to a correction \cite{Z,D} proportional to 
$r^{2}$) by
\begin{equation}
U_{C}(r) \approx U_{C}^{(b)}(r)-{\cal E},
\label{(6.2)}
\end{equation}
where $U_{C}^{(b)}(r)$ is the Coulomb potential for bare
uniformly charged nuclei ($R$ being the nuclear radius):
\begin{equation}
U_{C}^{(b)}(r)=(Z-2){e^{2}\over R} 
\left(3-{r^{2}\over R^{2}}\right), \; \; 
r \in [0,R[ , 
\label{(6.3)}
\end{equation}
\begin{equation}
U_{C}^{(b)}(r)=2(Z-2){e^{2}\over r}, \; \;
r \in ]R,\infty[,
\label{(6.4)}
\end{equation}
while ${\cal E}$ is the energy transferred to electrons. In 
non-metallic targets, $\cal E$ is the difference of electron
binding energies of the parent and daughter atoms. 
Eventually, upon defining
\begin{equation}
\kappa^{2} \equiv {2mE \over \hbar^{2}}, \; \;
\lambda \equiv l + {1 \over 2},
\label{(6.5)}
\end{equation}
\begin{equation}
v(r) \equiv {2m \over \hbar^{2}} U(r),
\label{(6.6)}
\end{equation}
stationary states are found by solving the stationary 
Schr\"{o}dinger equation
\begin{equation}
\left[{d^{2}\over dr^{2}}+\kappa^{2}
-{\left(\lambda^{2}-{1 \over 4}\right) \over r^{2}}
-v(r) \right]w_{\lambda}(\kappa;r)=0.
\label{(6.7)}
\end{equation}
 
The solutions of Eq. (6.7) are discussed in Ref. \cite{DZ}, but
here we would like to describe what new insight can be gained by
using the phase-integral method, following Ref. \cite{B4}. For this
purpose, we begin by remarking that, upon replacing $r$ with $z$,
Eq. (6.7) is of the form (2.1). The latter is solved by $\psi(z)$ 
having the form (2.2) provided that the exact phase integrand
$q(z)$ solves the differential equation
\begin{equation}
\chi(q(z)) \equiv q^{-{3 \over 2}}{d^{2}\over dz^{2}}q^{-{1 \over 2}}
+{R(z) \over q^{2}}-1=0,
\label{(6.8)}
\end{equation}
that is called the $q$-equation in Ref. \cite{B4}. Suppose now
that it is possible to determine a function 
$Q: z \rightarrow Q(z)$ that is an approximate solution of the
$q$-equation (6.8). This means that $\chi_{0}$, defined by
\begin{equation}
\chi_{0} \equiv \chi(Q(z))=Q^{-{3 \over 2}}
{d^{2}\over dz^{2}}Q^{-{1 \over 2}}
+{R(z)\over Q^{2}}-1,
\label{(6.9)}
\end{equation}
must be much smaller than $1$. The work in Ref. \cite{B4}
proves that the phase integrand $q(z)$ is related to the
base function $Q(z)$ by the asymptotic expansion
\begin{equation}
q(z) \sim Q(z) \sum_{n=0}^{N} Y_{2n},
\label{(6.10)}
\end{equation}
where, on defining the new independent variable (a sort of
approximate phase integral)
\begin{equation}
\zeta(z) \equiv \int^{z} Q(\tau)d\tau ,
\label{(6.11)}
\end{equation}
the first few $Y_{2n}$ functions are given explicitly 
by \cite{B4}
\begin{equation}
Y_{0}=1,
\label{(6.12)}
\end{equation}
\begin{equation}
Y_{2}={1 \over 2}\chi_{0},
\label{(6.13)}
\end{equation}
\begin{equation}
Y_{4}=-{1 \over 8} \left(\chi_{0}^{2}
+{d^{2}\over d \zeta^{2}}\chi_{0} \right),
\label{(6.14)}
\end{equation}
\begin{equation}
Y_{6}={1 \over 32} \left[2 \chi_{0}^{3}
+5 \left({d \chi_{0}\over d\zeta}\right)^{2}
+6 \chi_{0} {d^{2}\over d \zeta^{2}}\chi_{0}
+{d^{4}\over d\zeta^{4}}\chi_{0}\right].
\label{(6.15)}
\end{equation}
By virtue of Eqs. (6.1)-(6.4), the function $R(z)$ can
be written in the form
\begin{equation}
R(z)=-{\left(\lambda^{2}-{1 \over 4}\right)\over z^{2}}
+{a_{-1}\over z}+a_{0}+a_{1}z+{\rm O}(z^{2}),
\label{(6.16)}
\end{equation}
where the $a$'s are constants. Let us now assume that the
square of the freely specifiable base function is given by
\begin{equation}
Q^{2}(z)={b_{-2}\over z^{2}}+{b_{-1}\over z}
+b_{0}+b_{1}z+...,
\label{(6.17)}
\end{equation}
where the $b$'s are suitable constants. For the first-order
phase-integral approximation to be valid close to the origin,
one requires finiteness of the integral \cite{B4}
\begin{equation}
\mu(z,z_{0}) \equiv \left | \int_{z_{0}}^{z}
\Bigr | \chi_{0} Q(\tau) \Bigr | d\tau \right |
\label{(6.18)}
\end{equation}
as $z$ approaches $0$. After re-expressing $\chi_{0}$ in 
(6.9) in terms of $Q^{2}$ according to
\begin{equation}
\chi_{0}={1 \over 16Q^{6}}\left[5 
\left({dQ^{2}\over dz}\right)^{2}-4Q^{2}
{d^{2}\over dz^{2}}Q^{2}\right]
+{R(z) \over Q^{2}}-1,
\label{(6.19)}
\end{equation}
a patient calculation shows that \cite{B4}
\begin{eqnarray}
\chi_{0}Q &=& -{(\lambda^{2}+b_{-2})\over 
\sqrt{b_{-2}} \; z}
+{\left[\lambda^{2}
+b_{-2}-{1 \over 2} \right]b_{-1} \over
2 (b_{-2})^{3 \over 2}}
\nonumber \\
&+& {(a_{-1}-b_{-1}) \over \sqrt{b_{-2}}}
+{\rm O}(z).
\label{(6.20)}
\end{eqnarray}
Thus, finiteness of $\mu(z,z_{0})$ as $z$ approaches $0$ 
requires the elimination of the non-integrable term
proportional to ${1 \over z}$ in Eq. (6.20). This is
achieved if and only if
\begin{equation}
b_{-2}=-\lambda^{2},
\label{(6.21)}
\end{equation}
which implies in turn that
\begin{equation}
\lim_{z \to 0}z^{2}Q^{2}(z)=-\lambda^{2}
=-\left(l+{1 \over 2} \right)^{2},
\label{(6.22)}
\end{equation}
as well as \cite{B4}
\begin{equation}
\lim_{z \to 0} z^{2} \Bigr[Q^{2}(z)-R(z) \Bigr]
=-{1 \over 4}.
\label{(6.23)}
\end{equation}
The most convenient choice of $Q^{2}(z)$ in order to obtain
a stationary state that is regular at the origin at all
orders of approximation is \cite{B4}
\begin{equation}
Q^{2}(z)=R(z)-{1 \over 4z^{2}}.
\label{(6.24)}
\end{equation}

The advantage of the freely specifiable base function $Q(z)$
is that one has at disposal a new tool for finding approximate
forms of the stationary states as the potential (6.1) is
considered in greater detail, possibly including more
involved terms. The JWKB method does not have such a flexibility,
and higher orders of JWKB and phase-integral method may differ
in a substantial way \cite{B3,B4}.

We find it appropriate to end this section with an original 
calculation suggested by Eqs. (6.10)-(6.24). For this purpose, 
we assume to have chosen $Q^{2}(z)$ in the form (6.24), where
$R(z)$ takes the form (6.18) with vanishing ${\rm O}(z^{2})$ 
term (for simplicity). We then find the approximate phase integrand
$q(z)$ with arbitrary values of angular momentum quantum
number in the form
\begin{equation}
q(z) \sim \left(-{\lambda^{2}\over z^{2}}
+{a_{-1}\over z}+a_{0}+a_{1}z \right)^{1 \over 2}
\left(1+{\chi_{0}\over 2}\right),
\label{(6.25)}
\end{equation}
whrere, by virtue of (6.19) and (6.24),
\begin{eqnarray}
\; & \; & \chi_{0} = {1 \over 16} 
\left(-{\lambda^{2}\over z^{2}}+{a_{-1}\over z}
+a_{0}+a_{1}z \right)^{-3} 
\nonumber \\
& \times & \left[-{4 \lambda^{4}\over z^{6}}
+{12 \lambda^{2} a_{-1}\over z^{5}}
+{\Bigr(24 \lambda^{2}a_{0}-3(a_{-1})^{2}\Bigr) 
\over z^{4}} \right .
\nonumber \\
& + & \left . {(44 \lambda^{2} a_{1}-8a_{0}a_{-1}) \over z^{3}}
-{18 a_{1} a_{-1}\over z^{2}}+5 (a_{1})^{2} \right]
\nonumber \\
&+& {1 \over 4} \left(-\lambda^{2}+a_{-1}z+a_{0}z^{2}
+a_{1}z^{3}\right)^{-1}.
\label{(6.26)}
\end{eqnarray}
This formula yields in turn the asymptotic expansion of the
stationary state $\psi(z)$ by means of Eq. (2.4).

\section{Concluding remarks}
\setcounter{equation}{0}

Following the important findings in Refs.
\cite{Geiger,Gamow}, there has 
been valuable work on $\alpha$-decay for almost a century
by now \cite{Preston,Fermi,HW,Froman,Holstein,Z,D,AK,MANJU1,MANJU2,QI,DZ}. 
In particular, the work in Ref. \cite{Holstein} performs a very 
enjoyable presentation of four methods: complex eigenvalue, scattering
state method, semiclassical path integral, instanton method.
However, even the author of Ref. \cite{Holstein}, 
who was more familiar with the work
in Ref. \cite{Park}, was not aware of the one-directional
nature of connection formulae.   
Thus, our investigation is truly original, since it has applied
the work of Refs. \cite{B3,B4} to a nuclear physics problem
in which several generations of research workers were not aware 
of the proof of one-directional nature of connection formulae.  

Our original result (4.8) for the lowest $s$-wave metastable state
improves the values obtained in Refs. \cite{B2,Holstein}.
The authors of Ref. \cite{B2} find $\rho={3 \over 4} \pi$ because
they use in Eq. (3.8) the coefficients $A_{3}$ and $A_{4}$ enforced
by the wrong connection formulae (4.1). The work in Ref.
\cite{Holstein} finds instead $\rho=\pi$ because it approximates
the solutions of the equation 
$$
\tan (\rho)= - \nu
$$
by integer multiples of $\pi$. Moreover,
our formula (5.3) for the logarithm of the mean life of the 
radioactive nucleus yields a correction factor equal to $-1.025$
for the value obtained in Ref. \cite{B2}, and this prediction
can be checked against observation.

As far as we can see, our sources in the physics-oriented 
literature did an excellent work but
were misled by their lack of knowledge of one-directional nature
of connection formulae (cf. \cite{B3,B4}).
The main open problem is now
the application of the phase-integral perspective to the involved
models of modern nuclear physics. Our section $6$ has prepared
the ground for this purpose, describing in detail the logical
steps that are in order. Our original result for the approximate
form (6.25)-(6.26) of the phase integrand with arbitrary values
of the angular momentum quantum number provides,
as far as we can see, encouraging evidence
in favour of new tools being available for investigating
$\alpha$-decay from a phase-integral perspective.

\section*{Acknowledgments}
The author is grateful to the ``Ettore Pancini'' Physics
Department of Federico II University for hospitality and support. 

\begin{appendix}

\section{The $F$-matrix method}
\setcounter{equation}{0}

Let us assume that Eq. (2.1) is given, with the associated 
phase-integral functions (2.4). Following Ref. \cite{B4},
we consider the $a$-coefficients $a_{1}(z)$ and $a_{2}(z)$,
which are uniquely determined by the requirement that any
exact solution $\psi$ of Eq. (2.1) can be written in the form
\begin{equation}
\psi(z)=a_{1}(z)f_{1}(z)+a_{2}(z)f_{2}(z),
\label{(A1)}
\end{equation}
with first derivative given by 
\begin{equation}
{d \psi \over dz}=a_{1}(z){df_{1}\over dz}
+a_{2}(z){df_{2}\over dz}.
\label{(A2)}
\end{equation}
For Eq. (A2) to be satisfied, we have to impose that \cite{B4}
\begin{equation}
f_{1}(z){da_{1}\over dz}+f_{2}(z){da_{2}\over dz}=0.
\label{(A3)}
\end{equation}
Interestingly, Eq. (2.1) can be now replaced by a system of 
two coupled differential equations of first order, which can
be written in matrix form as \cite{B4}
\begin{equation}
{d \over dz} 
\left(\begin{matrix}
a_{1}(z) \\ a_{2}(z)
\end{matrix}\right)
=M(z) \left(\begin{matrix}
a_{1}(z) \\ a_{2}(z)
\end{matrix}\right),
\label{(A4)}
\end{equation}
having defined (see Eq. (2.9))
\begin{equation}
M(z)={i \over 2} \chi(z) q(z) 
\left(\begin{matrix}
1 & e^{-2iw(z)} \\
-e^{2i w(z)} & -1 
\end{matrix}\right).
\label{(A5)}
\end{equation}
Equation (A4) can be replaced by the integral equation
\begin{equation}
\left(\begin{matrix}
a_{1}(z) \\ a_{2}(z) 
\end{matrix}\right)
=\left(\begin{matrix}
a_{1}(z_{0}) \\ a_{2}(z_{0}) 
\end{matrix}\right)
+\int_{z_{0}}^{z}d\tau \; M(\tau)
\left(\begin{matrix}
a_{1}(\tau) \\ a_{2}(\tau) 
\end{matrix}\right),
\label{(A6)}
\end{equation}
whose solution can be obtained in closed form by an
iteration procedure that yields
\begin{equation}
\left(\begin{matrix}
a_{1}(z) \\ a_{2}(z) 
\end{matrix}\right) = F(z,z_{0})
\left(\begin{matrix}
a_{1}(z_{0}) \\ a_{2}(z_{0}) 
\end{matrix}\right),
\label{(A7)}
\end{equation}
where $F(z,z_{0})$ is a $2 \times 2$ matrix given by a
convergent series \cite{B4}. Such a matrix is the particular
solution of the differential equation 
\begin{equation}
{\partial \over \partial z}F(z,z_{0})=M(z)F(z,z_{0}),
\label{(A8)}
\end{equation}
that is equal to the $2 \times 2$ unit matrix for
$z=z_{0}$. The $F$-matrix satisfies the general relations
\cite{B4}
\begin{equation}
{\rm det} F(z,z_{0})=1,
\label{(A9)}
\end{equation}
\begin{equation}
F(z,z_{0})=F(z,z_{1})F(z_{1},z_{0}),
\label{(A10)}
\end{equation}
\begin{equation}
F(z_{0},z)=[F(z,z_{0})]^{-1}=
\left(\begin{matrix}
F_{22}(z,z_{0}) & -F_{12}(z,z_{0}) \\
-F_{21}(z,z_{0}) & F_{11}(z,z_{0})
\end{matrix}\right).
\label{(A11)}
\end{equation}
Useful estimates of the matrix elements of $F(z,z_{0})$ have
been obtained in Ref. \cite{B3} under the assumption that the
points $z$ and $z_{0}$ can be connected by a path in the complex
$z$-plane along which the absolute value of $e^{iw(z)}$ 
increases monotonically, in the non-strict sense, in the
direction from $z_{0}$ to $z$. Upon defining (cf. Eq. (2.10))
\begin{equation}
\mu = \mu(z,z_{0}) \equiv \left | 
\int_{z_{0}}^{z} |\chi(q(z')) \; q(z')|  dz' \right |,
\label{(A12)}
\end{equation}
these {\it basic estimates} read as \cite{B4}
\begin{equation}
\left | F_{11}(z,z_{0})-1 \right | \leq {\mu \over 2}
+ \; {\rm higher} \; {\rm powers} \; {\rm of} \; \mu ,
\label{(A13)}
\end{equation}
\begin{equation}
|F_{12}(z,z_{0})| \leq \left | e^{-2iw(z_{0})} \right |
\left({\mu \over 2}+ \; {\rm higher} \; {\rm powers} \;
{\rm of} \; \mu \right),
\label{(A14)}
\end{equation}
\begin{equation}
|F_{21}(z,z_{0})| \leq \left | e^{2iw(z_{0})} \right |
\left({\mu \over 2}+ \; {\rm higher} \; {\rm powers} \;
{\rm of} \; \mu \right),
\label{(A15)}
\end{equation}
\begin{equation}
|F_{22}(z,z_{0})-1| \leq {\mu \over 2}
+\left | e^{2i[w(z)-w(z_{0})]} \right |
\left({\mu^{2}\over 4}+ \; {\rm higher} \;
{\rm powers} \; {\rm of} \; \mu \right).
\label{(A16)}
\end{equation}
 
The parameters occurring in Eqs. (2.7) and (2.8) are real-valued
and can be defined as follows in terms of the $F$-matrix
\cite{B4}:
\begin{equation}
\alpha=\alpha(x_{1},x_{2})=|F_{11}(x_{1},x_{2})|,
\label{(A17)}
\end{equation}
\begin{equation}
\beta=\beta(x_{1},x_{2})=\pm {\rm arg} \; F_{11}(x_{1},x_{2}),
\label{(A18)}
\end{equation}
\begin{equation}
\gamma=\gamma(x_{1},x_{2})=
{\rm Re} \left[{F_{21}(x_{1},x_{2}) \over
F_{11}(x_{1},x_{2})} \right].
\label{(A19)}
\end{equation}
Strictly speaking, the possibility of writing $\gamma$ in
the form (A19) results from the simple but non-obvious
property, according to which \cite{B4}
$$
F_{21}(x_{1},x_{2}) F_{11}^{*}(x_{1},x_{2}) 
\mp {i \over 2}
$$
is real-valued.

\section{One-directional nature of the connection formula (2.18)}
\setcounter{equation}{0}

Suppose that, upon setting $d=1$ on the right-hand side of Eq.
(2.18), we are given a stationary state that, at a point 
$x_{2}$ of the classically allowed region, reads as
\begin{eqnarray}
\; & \; & \psi(x_{2})=2 \left | q^{-{1 \over 2}}(x_{2}) \right |
\; \cos \left[|w(x_{2})|-{\pi \over 4} \right] 
\nonumber \\
&=& a(x_{2}) \left | q^{-{1 \over 2}}(x_{2}) \right | 
e^{i |w(x_{2})|}
+b(x_{2}) \left | q^{-{1 \over 2}}(x_{2}) \right | 
e^{-i |w(x_{2})|},
\label{(B1)}
\end{eqnarray}
where $a(x_{2})=e^{-i{\pi \over 4}}$, $b(x_{2})=e^{i{\pi \over 4}}$.
The stationary state at a point $x_{1}$ in the classically forbidden
region reads therefore
\begin{equation}
\psi(x_{1})=c(x_{1}) \left|q^{-{1\over 2}}(x_{1})\right |
e^{|w(x_{1})|}+d(x_{1})
\left | q^{-{1\over 2}}(x_{1}) \right| e^{-|w(x_{1})|},
\label{(B2)}
\end{equation}
where the technique of Ref. \cite{B4} yields the formulae
(the approximate forms of the parameters $\alpha,\beta,\gamma$
being the ones given in our Eq. (2.11))
\begin{eqnarray}
c(x_{1})&=& \alpha e^{\left[-i\left({\pi \over 4}-\beta \right)\right]}
a(x_{2})+\alpha e^{\left[i \left({\pi \over 4}-\beta \right)\right]}
b(x_{2}) 
\nonumber \\
&=& 2 \alpha \sin \beta,
\label{(B3)}
\end{eqnarray}
\begin{eqnarray}
d(x_{1})&=& \left(\alpha \gamma+{i \over 2 \alpha} \right)
e^{-i \left({\pi \over 4}-\beta \right)} a(x_{2}) 
\nonumber \\
&+& \left(\alpha \gamma-{i \over 2 \alpha} \right)
e^{i \left({\pi \over 4}-\beta \right)} b(x_{2})
\nonumber \\
&=& {\cos \beta \over \alpha}
+2 \alpha \gamma \sin \beta.
\label{(B4)}
\end{eqnarray}
By virtue of Eqs. (B2)-(B4), one finds \cite{B4}
\begin{eqnarray}
\psi(x_{1})&=& \left | q^{-{1 \over 2}}(x_{1}) \right |
e^{-|w(x_{1})|}
\nonumber \\
& \times & \left \{ 2 \alpha \sin \beta e^{2 |w(x_{1})|}
+{\cos \beta \over \alpha}
+2 \alpha \gamma \sin \beta \right \} .
\label{(B5)}
\end{eqnarray}
By virtue of the approximate formulae (2.11) for 
$\alpha,\beta,\gamma$, the sum of terms within curly brackets
on the second line of (B5) can never approach $1$, 
and hence the stationary state in Eq. (B5) can never
approach the left-hand side of Eq. (2.18) 
with $c=0$ and $d=1$. Thus, the connection formula (2.18)
is one-directional \cite{B4}. 

As is stressed in Ref. \cite{B4}, the connection formula
has the same form for every order of the phase-integral
approximation.

\end{appendix}

\end{document}